\documentclass[sigconf]{acmart}

\usepackage{booktabs}

\usepackage{marvosym}

\AtBeginDocument{%
  \providecommand\BibTeX{{%
    \normalfont B\kern-0.5em{\scshape i\kern-0.25em b}\kern-0.8em\TeX}}}

\copyrightyear{2020} 
\acmYear{2020} 
\setcopyright{acmcopyright}\acmConference[ACSAC 2020]{Annual Computer Security
Applications Conference}{December 7--11, 2020}{Austin, USA}
\acmBooktitle{Annual Computer Security Applications Conference (ACSAC 2020),
December 7--11, 2020, Austin, USA}
\acmPrice{15.00}
\acmDOI{10.1145/3427228.3427274}
\acmISBN{978-1-4503-8858-0/20/12}

\begin{document}

\title{SEEF-ALDR: A Speaker Embedding Enhancement Framework via Adversarial Learning based Disentangled Representation}

\author{Jianwei Tai}
\affiliation{%
  \institution{Key Laboratory of Network Assessment Technology, Institute of Information Engineering, Chinese
Academy of Sciences; 
Beijing Key Laboratory of Network Security and Protection Technology, Institute of Information Engineering, Chinese
Academy of Sciences;
School of Cyber Security, University of Chinese Academy of Sciences;}
  \city{Beijing}
  \country{China}}
\email{taijianwei@iie.ac.cn}

\author{Xiaoqi Jia}
\authornote{\textsuperscript{\Letter}Xiaoqi Jia and \textsuperscript{\Letter}Weijuan Zhang are the corresponding authors.}
\affiliation{%
  \institution{Key Laboratory of Network Assessment Technology, Institute of Information Engineering, Chinese
Academy of Sciences; 
Beijing Key Laboratory of Network Security and Protection Technology, Institute of Information Engineering, Chinese
Academy of Sciences;
School of Cyber Security, University of Chinese Academy of Sciences;}
  \city{Beijing}
  \country{China}}
\email{jiaxiaoqi@iie.ac.cn}

\author{Qingjia Huang}
\affiliation{%
  \institution{Key Laboratory of Network Assessment Technology, Institute of Information Engineering, Chinese
Academy of Sciences; 
Beijing Key Laboratory of Network Security and Protection Technology, Institute of Information Engineering, Chinese
Academy of Sciences;
School of Cyber Security, University of Chinese Academy of Sciences;}
  \city{Beijing}
  \country{China}}
\email{huangqingjia@iie.ac.cn}

\author{Weijuan Zhang}
\authornotemark[1]
\affiliation{%
  \institution{Key Laboratory of Network Assessment Technology, Institute of Information Engineering, Chinese
Academy of Sciences; 
Beijing Key Laboratory of Network Security and Protection Technology, Institute of Information Engineering, Chinese
Academy of Sciences;
School of Cyber Security, University of Chinese Academy of Sciences;}
  \city{Beijing}
  \country{China}}
\email{zhangweijuan@iie.ac.cn}

\author{Haichao Du}
\affiliation{%
  \institution{Key Laboratory of Network Assessment Technology, Institute of Information Engineering, Chinese
Academy of Sciences; 
Beijing Key Laboratory of Network Security and Protection Technology, Institute of Information Engineering, Chinese
Academy of Sciences;
School of Cyber Security, University of Chinese Academy of Sciences;}
  \city{Beijing}
  \country{China}}
\email{Duhaichao@iie.ac.cn}

\author{Shengzhi Zhang}
\affiliation{%
  \institution{Department of Computer Science, Metropolitan College, Boston University}
  \city{Boston}
  \country{USA}}
\email{shengzhi@bu.edu}


\renewcommand{\shortauthors}{Jianwei and Xiaoqi, et al.}

\begin{abstract}
Speaker verification, as a biometric authentication mechanism, has been widely used due to the pervasiveness of voice control on smart devices. However, the task of ``in-the-wild'' speaker verification is still challenging, considering the speech samples may contain lots of identity-unrelated information, e.g., background noise, reverberation, emotion, etc. Previous works focus on optimizing the model to improve verification accuracy, without taking into account the elimination of the impact from the identity-unrelated information. To solve the above problem, we propose SEEF-ALDR, a novel Speaker Embedding Enhancement Framework via Adversarial Learning based Disentangled Representation, to reinforce the performance of existing models on speaker verification. The key idea is to retrieve as much speaker identity information as possible from the original speech, thus minimizing the impact of identity-unrelated information on the speaker verification task by using adversarial learning. 
Experimental results demonstrate that the proposed framework can significantly improve the performance of speaker verification by 20.3\% and 23.8\% on average over 13 tested baselines on dataset Voxceleb1 and 8 tested baselines on dataset Voxceleb2 respectively, without adjusting the structure or hyper-parameters of them. Furthermore, the ablation study was conducted to evaluate the contribution of each module in SEEF-ALDR. Finally, porting an existing model into the proposed framework is straightforward and cost-efficient, with very little effort from the model owners due to the modular design of the framework.
\end{abstract}



\maketitle

\section{Introduction}
Biometric authentication relies on biosensors to collect physiological and behavioral characteristics of users and applies the biostatistics principle to verify users' identity. Physiological features mainly include fingerprint, face, eyeball, iris, soundtrack, etc., and behavioral features include handwriting, gait, etc. As the advance of machine learning and the pervasiveness of mobile devices, biometric authentication has been widely used in everyone's daily life, e.g., fingerprint matching \cite{maltoni2005handbook}, face recognition \cite{phillips2000the}, iris recognition \cite{daugman2004how} and speaker verification \cite{nagrani2017voxceleb,chung2018voxceleb2,nagrani2020voxceleb} are supported by various smartphones these days. Recently, speaker verification has become a topic of interest for its applications in authentication \cite{chen2017you,voice-print}, forensics \cite{french2013an,gold2011international}, etc. For example, Barclays, a British banking company, has deployed voice authentication for its call centers since 2013 \cite{Barclays}. Alipay \footnote{Alipay is the largest mobile payment platform in the world, mainly used in China.} supports voice biometrics as an alternative authentication for users' convenience. Smart home devices like Amazon Echo, Apple HomePod, Google Home not only support voice control, but also integrate speaker verification to authenticate users for sensitive operations. The above application scenarios, with a large number of users worldwide, concern their private information as well as personal property.

The availability of ``in-the-wild'' datasets \cite{nagrani2017voxceleb,chung2018voxceleb2,nagrani2020voxceleb} helps speaker verification systems to become capable of dealing with real-world issues that require high accuracy and robustness in a variety of complex environments, e.g., grocery store, restaurant, etc. However, the task of ``in-the-wild'' speaker verification is still quite challenging. On one hand, the human voice may vary dramatically under different emotions and body situations, e.g., the physical condition of the throat and nose, etc. The diverse attributes of speech information hinder the extraction of speaker identity features. On the other hand, various environments may introduce different background noise and reverberation that can severely impact the speaker verification systems. Moreover, although the speech datasets are already large-scale compared with previous ones, they still cannot compensate for the variance introduced by the above-mentioned two factors, making it more difficult to learn uniform speaker embeddings. 

Recently, Convolutional Neural Network (CNN) has witnessed great success in the domain of face recognition and begins playing an important role in the domain of speaker verification on various speech datasets \cite{cai2018exploring,hajibabaei2018unified,bhattacharya2017deep,cai2018analysis,snyder2017deep,shon2018frame,snyder2018x}. For instance, previous works \cite{li2017deep,cai2018exploring,hajibabaei2018unified,xie2019utterance,liu2019large,yadav2018learning} have been proposed to further improve the performance of speaker verification. However, they directly process the original speech, which contains lots of irrelevant features other than the ones closely related to the speaker identity. Disentangled representation \cite{hsu2017unsupervised,yingzhen2018disentangled} is a straightforward idea to decouple the speaker identity features from the identity unrelated ones. Recently, disentangled representation has been explored in various computer vision applications, e.g., exposing invariant features for face recognition \cite{tran2017disentangled} and person re-identification \cite{ge2018fd}, attribute transfer via adversarial disentanglement \cite{liu2018exploring,lample2017fader}. In the audio domain, seminal works \cite{bhattacharya2019generative,zhou2019training,meng2019adversarial,tu2019variational} focus on robust identity representation by reducing the environmental complexity through adversarial learning. Nevertheless, they usually require additional and explicit supervision during the training process, which is not always feasible in real-world situations, e.g., background noise labels. Therefore, it is demanded that disentangled representation be effectively adopted for speaker embeddings to reduce the interference from identity-unrelated information.

To solve the above problem, we propose a novel speaker embedding enhancement framework via adversarial learning based disentangled representation, named SEEF-ALDR, which aims to minimize the impact of identity-unrelated information for speaker verification task. Inspired by recent advances in adversarial learning \cite{lample2017fader,liu2018exploring,hsu2019niesr}, we train SEEF-ALDR based on twin networks, with one extracting speaker identity features through a simple recognition training scheme and the other extracting identity-unrelated features through adversarial learning. The two sets of features obtained by the twin networks are later joined together in the reconstruction process that generates the reconstruction supervision signal. By optimizing the $L2$ distance between the reconstructed spectrogram and the original spectrogram, we ensure the union of the decoupled features still preserves the same information as the original spectrogram. 

With the speaker embeddings constructed from the speaker identity features, SEEF-ALDR significantly improves the performance of ``in-the-wild'' speaker verification. It is worth noting that SEEF-ALDR follows the modular design to ease the porting of existing speaker verification models. Experimental results show that SEEF-ALDR significantly improves the performance of the speaker verification task on ``in-the-wild'' datasets. The Equal Error Rate (EER) \cite{toh2008equal} is reduced by an average of 20.6\% on Voxceleb1 and 23.8\% on Voxceleb2 for all the tested baselines. To evaluate the contribution of each individual module, we conducted a comprehensive ablation study to learn the performance of each module introduced in the proposed framework for the speaker verification task. The experimental results show that each module in SEEF-ALDR collaborates effectively, thus achieving better speaker verification.

The contribution of this paper is highlighted as follows:
\begin{itemize}
\item (1) We propose a novel twin network-based speaker embedding enhancement framework to disentangle identity representation from the original speech by combining the autoencoder-like architecture and adversarial learning.

\item (2) SEEF-ALDR follows the modular design to facilitate the porting of existing speaker verification models, thus significantly improving the performance of all tested baselines for ``in-the-wild'' speaker verification without adjusting the structure or hyper-parameters of them.

\item (3) We only utilize speaker labels to train the eliminating encoder based on adversarial supervision to obtain the identity-unrelated information without explicit or manual labels.
\end{itemize}

The rest of this paper is organized as follows. Section \uppercase\expandafter{\romannumeral2} provides background and related works on speaker verification, disentangled representation, and speech recognition. In Section \uppercase\expandafter{\romannumeral3}, we detail the SEEF-ALDR approach proposed in this paper. In Section \uppercase\expandafter{\romannumeral4}, we conduct comprehensive experiments on the proposed framework to evaluate its performance on speaker verification and present the ablation study to demonstrate the contribution of each module in SEEF-ALDR. Section \uppercase\expandafter{\romannumeral5} discusses the potential to use the proposed framework for a novel identity impersonation attack and audio event detection task. Finally, we conclude in Section \uppercase\expandafter{\romannumeral6}.

\section{Background and Related Works}

\subsection{Speaker Verification}
\begin{figure}
  \centering
  \includegraphics[width=0.49\textwidth]{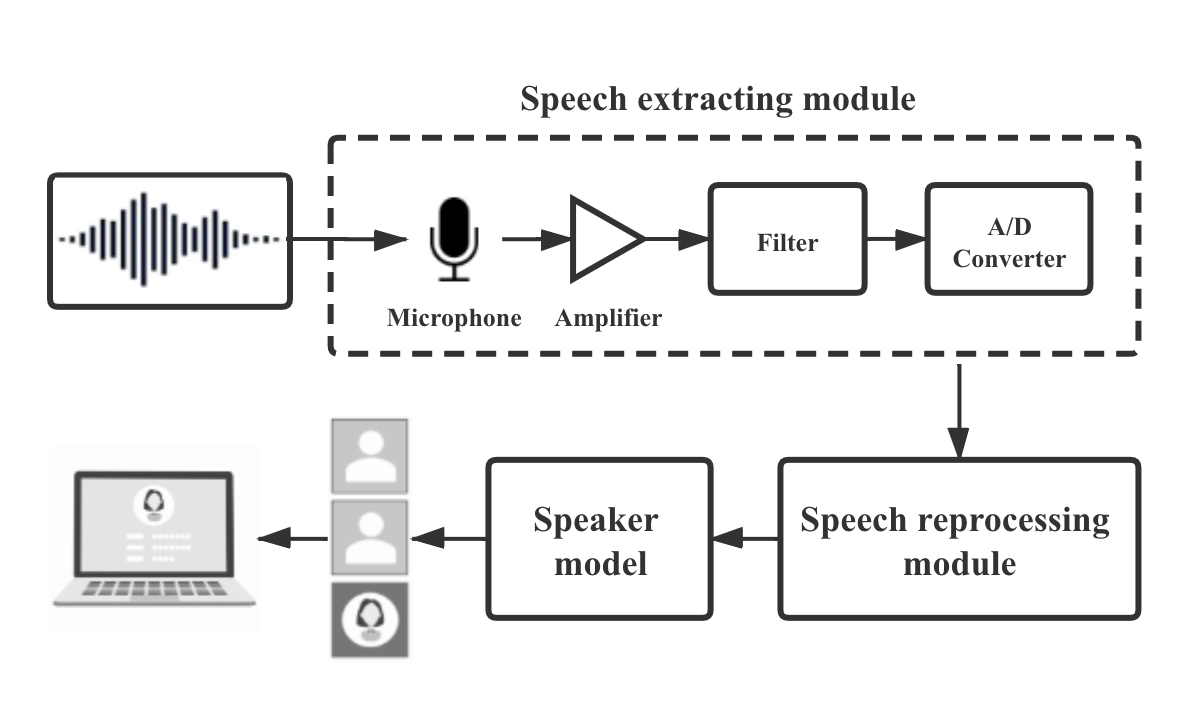} 
  \caption{The Architecture of Speaker Verification System based on Deep Learning.}
  \label{sv}
\end{figure}

According to the report \cite{TMR} provided by American consultancy Transparency Market Research, the global biometrics market will grow from 11.24 billion US dollars in 2015 to 23.3 billion US dollars in 2020, with a compound annual growth rate of 15.7\%. Among various biometric authentication techniques, speaker verification has become a hot topic with a promising development prospect.
Due to the advance of deep learning and artificial intelligence, recent studies \cite{nagrani2017voxceleb,chung2018voxceleb2,nagrani2020voxceleb,xie2019utterance,cai2018exploring,hajibabaei2018unified,li2017deep,bhattacharya2017deep,yadav2018learning} have introduced deep learning models into speaker verification tasks, thus significantly improving its accuracy. Therefore, the state-of-the-art speaker verification systems generally are built on top of deep learning models. Specifically, deep learning-based speaker verification system includes three major components: a speech extracting module, a speech pre-processing module, and a speaker model as shown in Figure \ref{sv}. Firstly, the speech extracting module collects the analog signal from a sound source using a microphone and digitizes the signal through A/D Converter. Then, the speech pre-processing module receives the digital signal and extracts the corresponding input vector according to different pre-processing methods, such as spectrogram \cite{hershey2016deep}, Mel Frequency Cepstrum Coefficient (MFCC) \cite{muda2010voice}, etc. Finally, the speaker model performs feature analysis on the input vector for speaker verification and determines whether the authentication is successful.

The availability of high-quality speech datasets \cite{nagrani2017voxceleb,chung2018voxceleb2} has greatly promoted the development of speaker verification. In terms of optimizing the model structure, for speaker recognition and verification, 2-dimensional CNN is often used for speaker recognition in both the time and frequency domain \cite{nagrani2017voxceleb,chung2018voxceleb2,cai2018exploring,hajibabaei2018unified,bhattacharya2017deep,cai2018analysis}, while 1-dimensional CNN is applied only in the time domain \cite{snyder2017deep,shon2018frame,snyder2018x}. The VGG and ResNet structures are proposed in \cite{nagrani2017voxceleb,chung2018voxceleb2} and in particular, Chung et al. \cite{chung2018voxceleb2} proposed to map voice spectrogram into a latent space to measure feature distances. Other works \cite{wan2018generalized,rahman2018attention} used LSTM-based front-end architectures to improve the accuracy of speaker feature extraction. Furthermore, utterance-level frame features are aggregated by a dictionary-based NetVLAD
or GhostVLAD layer for efficient speaker recognition \cite{xie2019utterance},  and Cai et al. \cite{cai2018exploring} obtained the utterance level representation by introducing a self-attentive pooling layer and the modified loss function. 

In terms of optimizing the objective function, although the softmax cross-entropy \cite{okabe2018attentive} is one of the most commonly used loss function to train speaker embedding models, it does not explicitly encourage discriminative learning. To alleviate this problem, the triplet loss \cite{li2017deep,zhang2017end} is introduced to optimize the accuracy of pre-trained models as a discriminative learning method. At the same time, Hajibabaei et al. \cite{hajibabaei2018unified} and Liu et al. \cite{liu2019large} systematically evaluate the impact of different loss functions and the presence or absence of the dropout method on the performance of the speaker embedding models. Moreover, Lee et al. \cite{lee2017discriminative} and LinWei-Wei et al. \cite{linwei-wei2018multisource} attempt to employ an autoencoder as its primary structure for speaker discrimination by acquiring speaker embeddings and has achieved remarkable performance improvements. However, despite the remarkable work \cite{kim2019deep} trying to minimize variability in speaker representation, identity-unrelated information has not been considered enough, which could suppress the performance of existing methods for speaker verification.

\subsection{Disentangled Representation}
To extract distinct and informative features from the original data, representation learning has received widespread attention in the artificial intelligence community. The autoencoder \cite{hinton2006reducing,hinton1994autoencoders} is proposed as a common representation learning method to extract feature representations from the original data. 
The encoder extracts feature representations from the original data, and then the decoder maps the embeddings from the feature space back to the input space, which are the processes of encoding and reconstruction. 
Due to the capability of encoding expressive information from the data space automatically, previous works \cite{vincent2008extracting,rifai2011contractive,kingma2013auto} developed various autoencoder-based models for different tasks, e.g., robust feature extraction, etc. Williams et al. \cite{williams2019disentangling} extracted different features through two different encoders with an autoencoder-like architecture for the conversion of speaking styles. Although existing works have made remarkable progress on various tasks, the problem of disentangling the feature representation space has not been studied comprehensively. 

Recently, learning disentangled representation based on generative models has attracted a lot of interest \cite{hsu2017unsupervised,yingzhen2018disentangled}. In the image domain, a series of works attempted to disentangle appropriate representations in various tasks, such as pose-invariant recognition \cite{tran2017disentangled} and identity-preserving image editing \cite{li2016deep,huang2017beyond}. In the audio domain, some studies also explored disentangled training by the adversarial supervision, such as using maximum mean discrepancy \cite{louizos2015variational} and GANs \cite{mathieu2016disentangling}. Subsequent works \cite{bhattacharya2019generative,zhou2019training,meng2019adversarial,tu2019variational} obtained robust speaker representations through domain adversarial learning by suppressing environmental complexity during speech recording. However, these works usually require explicit supervision during the training process and encode each attribute as a separate element in the feature vector, which may not be easily available in real-world situations. 
Hsu et al. \cite{hsu2019niesr} introduced an unsupervised method to induce the invariance of automatic speech recognition. From the perspective of data augmentation, Jati et al. \cite{jati2019multi} improved the performance of speaker verification through discriminative training. An inherent drawback of this approach is that the performance drops when dealing with complicated acoustic changes since it only learned very specific acoustic changes. Besides, data augmentation cannot explicitly ensure that irrelevant information is removed from the speaker representation, as shown by various detection tasks in \cite{raj2019probing}. Recently, Peri et al. \cite{peri2019robust} proposed an unsupervised adversarial invariance architecture to extract robust speaker-discriminative speech representations. However, this architecture does not generalize well, so it is difficult to port existing models to improve the performance of speaker verification.

\subsection{Speech Recognition}
Automatic Speech Recognition (ASR)\cite{hinton2012deep,graves2013speech} is to enables machines to recognize and understand human speech. Thanks to the significant improvement of the accuracy of ASR in recent years, voice has become an increasingly popular human-computer interaction mechanism due to its convenience and efficiency. Therefore, lots of voice-controllable systems and devices have been gradually introduced in the family life of the general public, e.g., Amazon Echo, Google Home, Apple HomePod, etc. Moreover,open-source platforms, such as Kaldi \cite{povey2011the}, Carnegie Mellon University's Sphinx \cite{huang1992the} and Mozilla DeepSpeech \cite{hannun2014deep} are also available for research community. The typical ASR system includes two main components: a speech acquisition module and a speech model. The former is composed of audio acquisition equipment and signal processing equipment. The latter consists of three sub-modules: feature extraction module, acoustic model, and language model. After the original audio passes through the power amplifier and filter, ASR system needs to extract acoustic features from the digitized audio signal. 

To a certain extent, the speaker verification system is similar to ASR system. Both of them are mainly composed of a speech acquisition module and a speech model, with the purpose of extracting and interpreting specific information from the input audio. However, ASR system needs to extract text-related information from the original speech, but the speaker verification system extracts the text-free, but identity-related information. In other words, ASR system concentrates on temporal sequence information, while the speaker verification system focuses on potential spatial information. Therefore, the advance of ASR system cannot be directly applied to the speaker verification system, and vice versa.

\section{Approach}
Our goal is to decouple the speaker identity information and identity-unrelated information contained in the original speech and obtain the purified speaker embedding to reduce the interference of the latter to the speaker verification task. In this section, we detail our SEEF-ALDR approach, a novel trainable network to learn the disentangled speaker identity features.

\begin{figure*}
  \centering
  \includegraphics[width=0.89\textwidth]{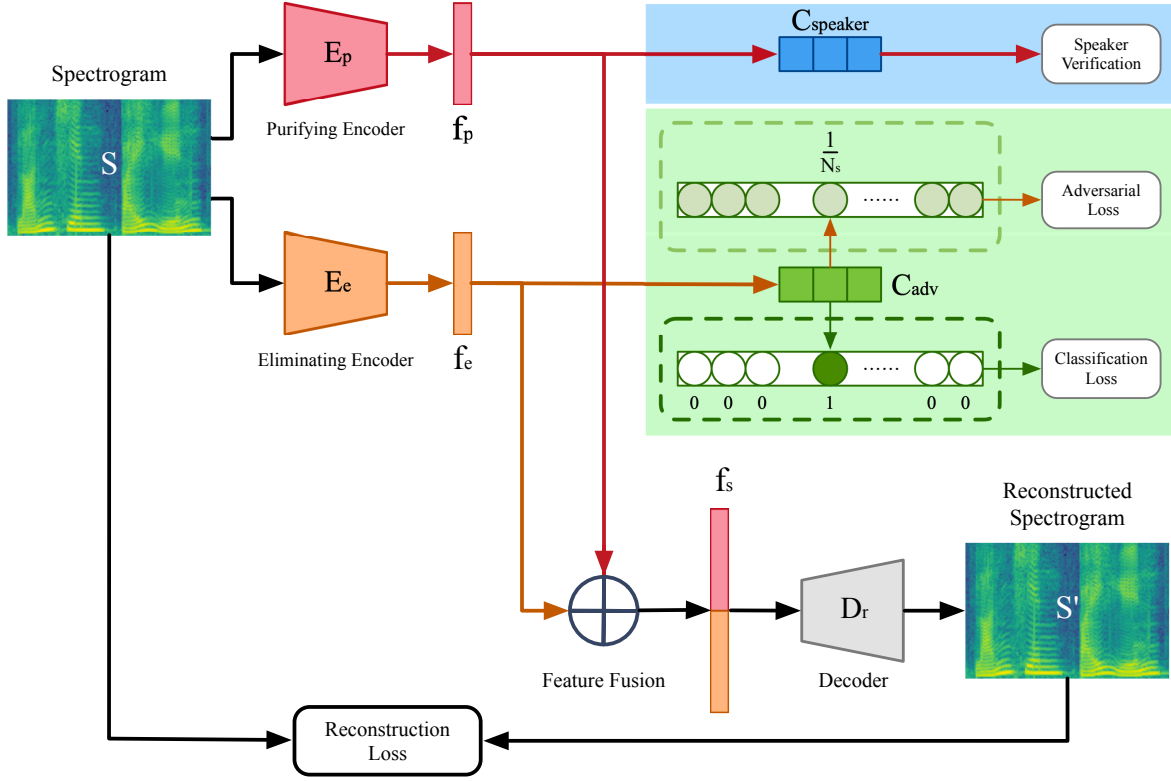} 
  \caption{The Architecture of SEEF-ALDR.}
  \label{pipeline}
\end{figure*}

\subsection{Overview}

The key idea of SEEF-ALDR is to decouple the identity-related information and identity-unrelated information from the input speech signal by using adversarial learning, and then extract the purified identity representation for the speaker verification task without interference from irrelevant features. Figure \ref{pipeline} shows the architecture of the proposed SEEF-ALDR. Given an input spectrogram $S$ from the original speech, the speaker purifying encoder $E_{p}$ extracts the identity-purified feature $f_{p}$, while the speaker eliminating encoder $E_{e}$ extracts the identity-unrelated feature $f_{e}$. $E_{p}$ and $E_{e}$ are trained based on twin networks, with the same multi-layer convolutional network structure as the to-be-integrated speaker verification models. The speaker identity labels act as a supervisory signal to train $E_{p}$, and also adversarially guide the training of $E_{e}$. 

Specifically, the eliminating encoder $E_{e}$ and adversarial classifier $C_{adv}$ perform a zero-sum game \cite{goodfellow2014generative}. $C_{adv}$ tries to classify the speaker's identity correctly based on the embeddings $f_{e}$ from $E_{e}$, while $E_{e}$ attempts to fool $C_{adv}$ by making the prediction results evenly distributed over each identity label, thus erasing the identity information from $f_{e}$. Finally, $f_{p}$ and $f_{e}$ are combined by the feature fusion to produce the fused spectrogram feature $f_{s}$. The decoder $D_{r}$ reconstructs a spectrogram $S'$ from $f_{s}$ to ensure that $S'$ is close enough to the original $S$. Such constraint on the spectrogram $S'$ requires the fusion feature $f_{s}$ to contain as much speech information as possible as the original $S$, so ideally feature $f_{p}$ and feature $f_{e}$ can be complementary. Therefore, as feature $f_{e}$ is encouraged to learn identity-unrelated information, feature $f_{p}$ will be forced to extract all identity-related information. 

By continuously iterating the training process, SEEF-ALDR can learn the distribution of the identity-related information and identity-unrelated information in the feature space from the speech samples and decouple them confidently, thereby obtaining more accurate speaker embeddings. Such embeddings generated by SEEF-ALDR can be used in the speaker verification task to further improve the performance of existing speaker verification models. In order to improve the portability of SEEF-ALDR, we follow a modular design. The encoders, classifiers, feature fusion, and the reconstruction decoder are all implemented in the form of modules. At the same time, all common pre-processing variables and device configurations are designed and implemented in the form of the hyper-parameters, which can be modified directly in the configuration file. Hence, we only need minor revision to the configuration file and the code of the to-be-integrated speaker verification models to port them into the encoders of our framework.

\subsection{Purifying Encoder}
The goal of the purifying encoder is to obtain a more accurate speaker embedding based on the identity-purified features extracted by the purifying encoder $E_{p}$, which can be simply written as $f_{p} = E_{p}\left ( S \right )$. Both the structure of $E_{p}$ and the objective function of the speaker classifier $C_{speaker}$ depend on the to-be-integrated speaker verification model (i.e., the baseline). The goal of $E_{p}$ can be simply understood as a multi-classification task. Generally, $softmax$ is often chosen to nonlinearly map the identity-purified features to the speaker prediction dimension $N_{s}$. Therefore, the prediction result of $y_{p}$ can be written as:
\begin{eqnarray}
y_{p} = \mathrm{softmax} \left (C_{speaker}\left (E_{p}\left ( S \right ) \right ) \right ) . 
\end{eqnarray}

We compare $y_{p}$ and the encoded speaker identity distribution $p_{s}$ by cross entropy. The objective function $L_{p}^{S}$ of training the speaker classifier $C_{speaker}$ with $softmax$ can be expressed as:
\begin{eqnarray}
\label{formula2}
L_{p}^{S} = -\sum_{j=1}^{N_{s}} p^{j}_{s} \log \left ( y^{j}_{p} \right ) .
\end{eqnarray}

To improve the performance of speaker verification, previous works like $A-softmax$ \cite{liu2017sphereface,pellegrini2019cosine} have been proposed based on the standard $softmax$ function. We also reproduced $A-softmax$, which guarantees that the angle of different classes in the feature space is as large as possible. In other words, the angular margin between each input sample and the center of its correct category is dictated as $m$ times smaller than those between it and the centers of other wrong categories. We define $f\left ( x_{j} \right )$ as the corresponding output of the penultimate layer of speaker classifier $C_{speaker}$ for the input $x_{j}$, and $t_{j}$ as the corresponding target label. When the loss function $A-softmax$ is used, the prediction score given by $A-softmax$ is the weighted average over the cosine similarity with weight $\lambda _{cos}$ that decreases during the training process, which can be expressed as:
\begin{eqnarray}
f_{t_{j}} = \frac{ \lambda _{cos} \left \| f\left ( x_{j} \right ) \right \| \cos \left ( \theta _{t_{j}} \right )  +   \left \| f\left ( x_{j} \right ) \right \| \phi  \left ( \theta _{t_{j}} \right ) }{\lambda _{cos} + 1}  
\end{eqnarray}
with $\phi  \left ( \theta _{t_{j}} \right ) = \left ( -1 \right )^{k} \cos \left ( m \theta _{t_{j}} \right ) - 2k$, $\theta _{t_{j}} \in \left [ \frac{k\pi }{m} , \frac{ \left ( k+1 \right ) \pi }{m}  \right ]$ and $k \in \left [ 0,m-1 \right ]$, where $m$ is an integer to control the size of angular margin. Therefore, the objective function to train the speaker classifier $C_{speaker}$ can be written as:
\begin{eqnarray}
\label{formula4}
L_{p}^{A-S} = -\sum_{j=1}^{N_{s}} p^{j}_{s} \log \left (   \frac{ e^{f_{t_{j}}} }{ e^{f_{t_{j}}} + \sum _{i \neq t_{j}} e^{\left \| x_{j} \right \| \cos \left ( \theta _{i} \right )} }    \right ). 
\end{eqnarray}

With a distinct geometric interpretation, the encoder supervised by $A-softmax$ can learn features that construct a discriminative angular distance metric on the hypersphere manifold \cite{liu2017sphereface}. When $m \geq 2$, $A-softmax$ loss produces more accurate classification by forcing different categories to have certain margins of angular classification in the feature space. To simplify the expression, we use $L_{p}$ to indicate the objective function to train the speaker classifier $C_{speaker}$ thereafter in the paper, which varies based on different implementation. 


\subsection{Eliminating Encoder}
The adversarial classifier $C_{adv}$ is to classify the speaker identity based on the predicted distribution $y_{e} = softmax\left ( C_{adv}\left (  E_{e}\left ( S \right ) \right ) \right )$. The eliminating encoder $E_{e}$ is trained to fool $C_{adv}$ through an adversarial supervision signal, so that $C_{adv}$ outputs the same probability over each prediction class. Hence, $E_{e}$ can successfully extract the identity-unrelated features $f_{e} = E_{e}\left (S \right )$.
The objective function $L^{adv}_{s}$ of $C^{adv}$ directs speaker identification based on feature $f_{e}$, and is constrained through the cross entropy loss as below:
\begin{eqnarray}
\label{formula5}
L^{adv}_{s} = -\sum_{j=1}^{N_{s}} t^{j}_{s} \log \left ( y^{j}_{e} \right ) 
\end{eqnarray}
where $t_{s}$ means the ground truth of speakers' identity. Note that the gradient of $L^{adv}_{s}$ only propagates back to $C_{adv}$, without updating any layer of $E_{e}$. 

The eliminating encoder $E_{e}$ should be trained to fool $C^{adv}$, so the speaker identity distribution $u_{s}$ is set as a constant for each speaker label, equal to $\frac{1}{N_{s}}$ in the cross-entropy loss of $softmax$. The objective function $L^{adv}_{e}$ of $E_{e}$ can be written as follow:
\begin{eqnarray}
\label{formula6}
L^{adv}_{e} = \sum_{j=1}^{N_{s}} u^{j}_{s} \log \left ( y^{j}_{e} \right ) =\frac{1}{N_{s}}\sum_{j=1}^{N_{s}} \log \left ( y^{j}_{e} \right ) .
\end{eqnarray}

The gradient of $L^{adv}_{e}$ only propagates back to $E_{e}$, rather than $C_{adv}$. If we allow the gradient of $L^{adv}_{e}$ to update $C_{adv}$, the encoder $E_{e}$ can easily cheat $C_{adv}$, e.g, by only revising the classifier $C_{adv}$ to produce non-informative output.  
Hence, the encoder $E_{e}$ cannot ensure that feature $f_{e}$ will extract the identity-unrelated information under these circumstances. Therefore, by combining both $L^{adv}_{e}$ and $L^{adv}_{s}$, our framework can leverage the advantages of each of them and be coordinated to work together towards the identity-unrelated feature through disentangled speaker information. Benefiting from the advantages of adversarial learning, the speaker eliminating encoder $E_{e}$ can learn irrelevant features as accurately as possible, which in turn guarantees the correctness of speaker identity features. 


\subsection{Reconstruction Decoder}
Ideally, the combination of the decoupled features $f_{p}$ and the identity-unrelated features $f_{e}$ should be exactly the representation of the input spectrogram $S$. Hence, defining the feature fusion module $\bigoplus$ based on the concatenating operation, we fuse $f_{p}$ and $f_{e}$ together into a complete feature $f_{s} = f_{p} \bigoplus f_{e}$, and make the decoder $D_{r}$ reconstruct the spectrogram $S'=D_{r} \left ( f_{s} \right )$. To simply measure the difference between the reconstructed spectrogram $S'$ and the original spectrogram $S$, we utilize $l_{2}$ distance\footnote{Since the input spectrogram is in the form of a 2-dimensional matrix, $l_{2}$ distance is a commonly used measure to compare the similarity between two 2-dimensional matrices.} to define the reconstruction loss $L_{r}$ as below:
\begin{eqnarray}
\label{formula7}
L_{r}=\frac{1}{2} \left \|  D_{r} \left ( f_{p} \bigoplus  f_{e}  \right ) - S  \right \| ^{2}_{2} .
\end{eqnarray}

As mentioned above, the adversarial supervision signal encourages the eliminating encoder $E_{e}$ to extract identity-unrelated features. In contrast, the reconstruction loss guides the purifying encoder $E_{p}$ to embed the remaining identity-purified features by the fidelity of the spectrogram reconstruction, i.e., ensuring $f_{s}$ contains the complete representation from the original spectrogram $S$. In general, the purifying encoder $E_{p}$ needs to be trained first to reach a certain level of accuracy in the task of speaker verification. Then the $E_{e}$ initiates its networks by inheriting the weights from $E_{p}$ and begins the adversarial learning. Meanwhile, the reconstruction decoder takes $f_{s}$ from the feature fusion, and begins the process of spectrogram reconstruction to interactively train $E_{p}$ and $E_{e}$ for obtaining complementary feature pairs, i.e., the gradient of the reconstruction loss propagates back to the encoder $E_{p}$ and the encoder $E_{e}$. 


\subsection{Objective Function}
Learning the speaker identity representation involves multiple objectives that consist of the feature extraction loss $L_{p}$, the adversarial losses $L^{adv}_{s}$ and $L^{adv}_{e}$, as well as the speech reconstruction loss $L_{r}$. Therefore, the overall objective function of SEEF-ALDR with a weighted combination of them is as below:
\begin{eqnarray}
\label{objective}
L = \lambda _{p} L_{p} + \lambda _{adv} \left (  L^{adv}_{s} + L^{adv}_{e}  \right ) + \lambda _{r} L_{r} 
\end{eqnarray}
where $\lambda _{p}$, $\lambda _{adv}$ and $\lambda _{r}$ are weight parameters. $L_{p}$ ($L^{S}_{p}$ for $softmax$ and $L^{A-S}_{p}$ for $A-softmax$), $L^{adv}_{s}$, $L^{adv}_{e}$ and $L_{r}$ can be referred to in Equations \ref{formula2}, \ref{formula4}, \ref{formula5}, \ref{formula6}, \ref{formula7}, respectively. To minimize the overall loss, the stochastic gradient descent solver is used to solve Equation \ref{objective}. Since SEEF-ALDR is composed of several modules, in our experiments, we provide an ablation study to evaluate the contribution of each module.

\begin{table}
\caption{Training and Testing Dataset on Voxceleb1.}
\label{split1}
\centering
\begin{tabular}{lccc}
\toprule
Dataset  &   Training   &  Testing  & Total \\
\midrule
\#POIs & 1211 & 40 & 1251 \\
\#Utterances & 148,642 & 4,874 & 153,516 \\ 
\bottomrule
\end{tabular}
\end{table}

\begin{table}
\caption{Training and Testing Dataset on Voxceleb2.}
\label{split2}
\centering
\begin{tabular}{lccc}
\toprule
 Dataset  &   Training   &  Testing  & Total \\ 
\midrule
\#POIs & 5994 & 118 & 6112 \\
\#Utterances & 1,092,009 & 36,237 & 1,128,246 \\ 
\bottomrule
\end{tabular}
\end{table}

\section{Experiments}
In this section, we first elaborate on the experimental setup, and then present the results of SEEF-ALDR on enhancing the speaker verification of existing models. Finally, we conduct the ablation study to evaluate the contribution of each module in SEEF-ALDR.

\subsection{Experimental Setup}
\noindent\textbf{Datasets.}
Datasets Voxceleb1 \cite{nagrani2017voxceleb} and Voxceleb2 \cite{chung2018voxceleb2} are large-scale, text-independent speech databases collected from videos on YouTube, and can be used for both speaker identification and verification tasks. In our experiments, we only use audio files from them for speaker verification task. Voxceleb1 contains 153,516 utterances from 1,251 speakers and Voxceleb2 contains 1,128,246 utterances from 6,112 speakers. Both of them are fairly gender-balanced, with 45\% of female speakers for Voxceleb1 and 39\% of female speakers for Voxceleb2. The speakers span a wide range of races, professions, ages, emotions, accents, etc. The source video contained in the dataset was recorded in quite diverse visual and auditory environments, including interviews from the red carpet, outdoor sports fields as well as quiet indoor studios, public speeches to lots of audiences, etc. Therefore, these ``in-the-wild'' speech samples contain a large amount of identity-unrelated information and noise. Experiments on these samples help highlight the advance of SEEF-ALDR in disentangled representation. Table \ref{split1} and Table  \ref{split2} show the division of training and testing data of Voxceleb1 and Voxceleb2. It is worth mentioning that there are no overlapping identities between the training dataset of VoxCeleb2 and the overall dataset VoxCeleb1. We train the proposed framework SEEF-ALDR on datasets Voxceleb1 and Voxceleb2.

\noindent\textbf{Network Architecture}
SEEF-ALDR consists of five components: the speaker purifying encoder $E_{p}$, the speaker eliminating encoder $E_{e}$, the speaker classifier $C_{speaker}$, the adversarial classifier $C_{adv}$, and the reconstruction decoder $D_{r}$. The network architecture of the speaker purifying encoder $E_{p}$ in SEEF-ALDR depends on the structure of to-be-integrated speaker verification models, such as VGGnet, Resnet34, Resnet50, etc. We intend to set the architecture of the speaker eliminating encoder $E_{e}$ the same as $E_{p}$ to make it easier to train the entire framework, though it can be totally different. For instance, the backbone of $E_{p}$ and $E_{e}$ is based on ResNet-34 when we port \cite{he2016deep} (using ResNet-34) into our framework, appended with the global temporal pool (TAP) layer to embed variable-length input speech into the fixed-length speaker feature. Furthermore, to reproduce the baseline models, we introduce another self-attentive pooling (SAP) layer based on \cite{okabe2018attentive}. There is one fully connected layer in $C_{speaker}$, and three convolutional layers as well as three fully connected layers included in $C_{adv}$. A simple implementation of $D_{r}$ is composed of three fully connected layers and ten fractionally-strided convolution layers \cite{radford2016unsupervised} interlaced with batch normalization layers to obtain the reconstructed spectrogram. To further ensure the efficiency and stability of the training process, the decoder is also consistent with the corresponding encoder to do upsampling by using deconvolution (or called transposed convolution).

\noindent\textbf{Initialization}
SEEF-ALDR is trained on a Linux server with i7-8700K CPU, 32GB memory, and three NVIDIA Titan V GPUS connected in an end-to-end manner. During pre-processing, spectrogram of all the input speech samples are extracted in a sliding window fashion using a hamming window \cite{nagrani2017voxceleb,chung2018voxceleb2} with $width = 25ms$ and $step = 10ms$, and normalized to a standardized variable (with mean of 0 and standard deviation of 1). Since the duration of the speech samples is different, we randomly choose a three-second temporal segments from each spectrogram to ensure that the size of the training samples is consistent. The batch size of the input speech is set as 64 and the model is trained through SGD optimizer with $momentum = 0.9$ and $weight\_decay = 5e-4$. The initial learning rate is set as $10^{-2}$, and reduced by 10\% per cycle of the previous learning rate until $10^{-6}$. When $A-softmax$ is used as the loss function of the speaker classifier, the angular margin factor is set as $m = 4$, and the weight factor is set as $\lambda _{cos} = 5$. The weight parameters in the training process are set as $\lambda _{p} = 1$ for $L_{p}$, $\lambda _{r} = 0.02$ for $L_{r}$, and $\lambda _{adv} = 0.1$ for $L^{adv}_{s}$ and $L^{adv}_{e}$ respectively.
The training balance between the eliminating encoder and the adversarial classifier needs to be adjusted empirically to improve the accuracy of feature decoupling.


\begin{table*}
\caption{Speaker Verification Performance of SEEF-ALDR Trained on Voxceleb1. EER-IP represents the EER Improvement over the Baseline.}
\label{vox1}
\centering
\begin{tabular}{lcccccccc}

\toprule
 & \ Model \ & Loss Function  & Dims & Aggregation & Metric & EER (\%) & $C_{det}$ & \ EER-IP \\ 
\midrule
Nagrani et al. \cite{nagrani2017voxceleb} & VGG-M & Softmax & 512 & TAP & Cosine & 7.8 & 0.71 & - \\
\textbf{SEEF-ALDR} & VGG-M & Softmax & 512 & TAP & Cosine & \textbf{6.51} & \textbf{0.619} & \textbf{16.7\%} \\ 
\midrule
Li et al. \cite{li2017deep} & ResCNN & Softmax+Triplet & 512 & TAP & Cosine & 4.80 & N/A & - \\
\textbf{SEEF-ALDR} & ResCNN & Softmax+Triplet & 512 & TAP & Cosine & \textbf{4.02} & \textbf{0.469} & \textbf{16.3\%} \\ 
\midrule
Bhattacharya et al. \cite{bhattacharya2017deep} & VGGnet & Softmax & 512 & TAP & PLDA & 4.52 & N/A & - \\
\textbf{SEEF-ALDR} & VGGnet & Softmax & 512 & TAP & PLDA & \textbf{3.95} & \textbf{0.439} & \textbf{12.6\%} \\ 
\midrule
Cai et al. \cite{cai2018exploring} & ResNet-34 & Softmax & N/A & TAP & Cosine & 5.48 & 0.553 & - \\
\textbf{SEEF-ALDR} & ResNet-34 & Softmax & 256 & TAP & Cosine & \textbf{4.31} & \textbf{0.454} & \textbf{21.4\%} \\
Cai et al. \cite{cai2018exploring} & ResNet-34 & Softmax & N/A & TAP & PLDA & 5.21 & 0.545 & - \\
\textbf{SEEF-ALDR} & ResNet-34 & Softmax & 256 & TAP & PLDA & \textbf{4.35} & \textbf{0.479} & \textbf{16.5\%} \\
Cai et al. \cite{cai2018exploring} & ResNet-34 & A-Softmax & N/A & TAP & Cosine & 5.27 & 0.439 & - \\
\textbf{SEEF-ALDR} & ResNet-34 & A-Softmax & 256 & TAP & Cosine & \textbf{4.26} & \textbf{0.433} & \textbf{19.2\%} \\ 
\midrule
Cai et al. \cite{cai2018exploring} & ResNet-50 & Softmax & N/A & SAP & Cosine & 5.51 & 0.522 & - \\
\textbf{SEEF-ALDR} & ResNet-50 & Softmax & 256 & SAP & Cosine & \textbf{4.40} & \textbf{0.469} & \textbf{20.1\%} \\
Cai et al. \cite{cai2018exploring} & ResNet-50 & A-Softmax & N/A & SAP & Cosine & 4.90 & 0.509 & - \\
\textbf{SEEF-ALDR} & ResNet-50 & A-Softmax & 256 & SAP & Cosine & \textbf{4.08} & \textbf{0.455} & \textbf{16.7\%} \\
\midrule
Hajibabaei et al. \cite{hajibabaei2018unified} & ResNet-20 & Softmax & 256 & TAP & Cosine & 6.98 & 0.540 & - \\
\textbf{SEEF-ALDR} & ResNet-20 & Softmax & 256 & TAP & Cosine & \textbf{4.56} & \textbf{0.503} & \textbf{34.7\%} \\
Hajibabaei et al. \cite{hajibabaei2018unified} & ResNet-20 & Softmax & 128 & TAP & Cosine & 6.73 & 0.526 & - \\
\textbf{SEEF-ALDR} & ResNet-20 & Softmax & 128 & TAP & Cosine & \textbf{4.48} & \textbf{0.497} & \textbf{33.4\%} \\ 
Hajibabaei et al. \cite{hajibabaei2018unified} & ResNet-20 & Softmax & 64 & TAP & Cosine & 6.31 & 0.527 & - \\
\textbf{SEEF-ALDR} & ResNet-20 & Softmax & 64 & TAP & Cosine & \textbf{4.37} & \textbf{0.494} & \textbf{30.7\%} \\ 
\midrule
Hajibabaei et al. \cite{hajibabaei2018unified} & ResNet-20 & A-Softmax & 128 & TAP & Cosine & 4.40 & 0.451 & - \\
\textbf{SEEF-ALDR} & ResNet-20 & A-Softmax & 128 & TAP & Cosine & \textbf{3.81} & \textbf{0.437} & \textbf{13.4\%} \\
Hajibabaei et al. \cite{hajibabaei2018unified} & ResNet-20 & A-Softmax & 64 & TAP & Cosine & 4.29 & 0.442 & - \\
\textbf{SEEF-ALDR} & ResNet-20 & A-Softmax & 64 & TAP & Cosine & \textbf{3.62} & \textbf{0.437} & \textbf{15.6\%} \\
\bottomrule

\end{tabular}


\end{table*}

\subsection{Performance of SEEF-ALDR}
We selected several state-of-the-art speaker verification models as representatives to port into SEEF-ALDR and evaluate the effectiveness of SEEF-ALDR \cite{nagrani2017voxceleb,chung2018voxceleb2,xie2019utterance,cai2018exploring,hajibabaei2018unified,li2017deep,bhattacharya2017deep} (the baseline performance). When reproducing those models, we ensure that the model structure, loss function, test dataset, and similarity metric are consistent with those in the original paper. After porting them into SEEF-ALDR, we retrain SEEF-ALDR on Voxceleb1 and Voxceleb2 respectively. We choose two metrics: the detection cost function ($C_{det}$) \cite{greenberg2013the} and the Equal Error Rate (EER) \cite{toh2008equal}, to evaluate the performance of SEEF-ALDR on the speaker verification task. $C_{det}$ can be computed as below:
\begin{eqnarray}
C_{det} = C_{miss} P_{miss} \cdot P_{tar} + C_{fa} P_{fa} \cdot \left ( 1-P_{tar}  \right )
\end{eqnarray}
where $P_{miss}$ is the probability of the miss and $P_{fa}$ is the probability of the false alarm. The prior target probability $P_{tar}$ is set as 0.01, and both the cost of a miss $C_{miss}$ and the cost of a false alarm $C_{fa}$ have equal weight parameter of 1.0. We demonstrate the effectiveness of SEEF-ALDR by showing the improvement of EER after porting the baseline into SEEF-ALDR, i.e., the percentage of reduction in EER. 

\begin{table*}
\caption{Speaker Verification Performance of SEEF-ALDR trained on Voxceleb2. EER-IP represents the EER Improvement over the Baseline.}
\label{vox2}
\centering
\begin{tabular}{lccccccc}

\toprule
 & \ Model \ & \ \ Loss Function \ \  & \ Test set \ & EER (\%) & $C_{det}$  & EER-IP \\ 
\midrule

Xie et al. \cite{xie2019utterance} & Thin ResNet+NV & Softmax & Voxceleb1 & 3.57 & N/A & - \\
\textbf{SEEF-ALDR} & Thin ResNet+NV & Softmax & Voxceleb1 & \textbf{2.85} & \textbf{0.327} & \textbf{20.2\%} \\ 
Xie et al. \cite{xie2019utterance} & Thin ResNet+GV & Softmax & Voxceleb1 & 3.22 & N/A & - \\
\textbf{SEEF-ALDR} & Thin ResNet+GV & Softmax & Voxceleb1 & \textbf{2.61} & \textbf{0.335} & \textbf{19.0\%} \\  
 
\midrule
Xie et al. \cite{xie2019utterance} & Thin ResNet+NV & Softmax & Voxceleb1-E & 3.24 & N/A & - \\
\textbf{SEEF-ALDR} & Thin ResNet+NV & Softmax & Voxceleb1-E & \textbf{2.87} & \textbf{0.373} & \textbf{11.4\%} \\ 
Xie et al. \cite{xie2019utterance} & Thin ResNet+GV & Softmax & Voxceleb1-H & 5.17 & N/A & - \\
\textbf{SEEF-ALDR} & Thin ResNet+GV & Softmax & Voxceleb1-H & \textbf{4.52} & \textbf{0.520} & \textbf{12.6\%} \\
\midrule
\midrule
Chung et al. \cite{chung2018voxceleb2} & ResNet-34 & Softmax + Contrastive & Voxceleb1 & 5.04 & 0.543 & - \\
\textbf{SEEF-ALDR} & ResNet-34 & Softmax + Contrastive & Voxceleb1 & \textbf{3.08} & \textbf{0.334} & \textbf{38.8\%}\\ 
Chung et al. \cite{chung2018voxceleb2} & ResNet-50 & Softmax + Contrastive & Voxceleb1 & 4.19 & 0.449 & - \\
\textbf{SEEF-ALDR} & ResNet-50 & Softmax + Contrastive & Voxceleb1 & \textbf{2.75} & \textbf{0.326} & \textbf{34.4\%} \\ 

\midrule
Chung et al. \cite{chung2018voxceleb2} & ResNet-50 & Softmax + Contrastive & Voxceleb1-E & 4.42 & 0.524 & - \\
\textbf{SEEF-ALDR} & ResNet-50 & Softmax + Contrastive & Voxceleb1-E & \textbf{3.25} & \textbf{0.398} & \textbf{26.5\%} \\ 
Chung et al. \cite{chung2018voxceleb2} & ResNet-50 & Softmax + Contrastive & Voxceleb1-H & 7.33 & 0.673 & - \\
\textbf{SEEF-ALDR} & ResNet-50 & Softmax + Contrastive & Voxceleb1-H & \textbf{5.30} & \textbf{0.575} & \textbf{27.7\%} \\

\bottomrule

\end{tabular}


\end{table*}

We first utilize the training dataset and testing dataset from Voxceleb1 for speaker verification. To calculate speaker similarity for the verification task, we choose the common cosine distance as the similarity matrix. In addition to the cosine distance, probabilistic linear discriminant analysis (PLDA) \cite{prince2007probabilistic} is adopted in the state-of-the-art speaker verification system \cite{cai2018exploring,bhattacharya2017deep} to measure the distance between two speaker embeddings. As shown in Table \ref{vox1}, the experimental results demonstrate that SEEF-ALDR can significantly improve the performance of each baseline, with an average improvement of 20.6\% on EER. 

The best EER improvement in Table \ref{vox1} comes from \cite{hajibabaei2018unified} with ResNet-20 as the baseline model, all above 30\%. In contrast, also using $softmax$, TAP and Cosine, 16.7\% improvement over VGG-M \cite{nagrani2017voxceleb} is observed. The reason for such a significant difference is that the network structure of VGG-M is much simpler, compared with that of ResNet-20, so the potential for performance improvement is also limited. Overall, the lowest two EER improvements are 12.6\% \cite{bhattacharya2017deep} and 13.4\% \cite{hajibabaei2018unified}, where $A-Softmax$ or PLDA are used to replace $softmax$ or Cosine accordingly in their baseline models. The possible reason for such low improvement compared with others might be that those models have gone through task-specific optimization, thus not leaving too much room for SEEF-ALDR to further improve the performance of speaker verification. Overall, the experimental results already demonstrate that the proposed SEEF-ALDR can significantly enhance the capability of the baseline models to learn the identity-related representation for better speaker verification.

To verify the performance of SEEF-ALDR on a larger dataset, we choose the training set from Voxceleb2 and then choose three different testing sets from Voxceleb1 and Voxceleb2: original $Voxceleb1$ test set, new $Voxceleb1-H$ test set and new $Voxceleb1-E$ test set \cite{chung2018voxceleb2}. Due to the limited number of speakers in the original $Voxceleb1$ test set (only 40 speakers), a good performance on it does not necessarily indicate the model trained using the larger training set from Voxceleb2 can perform well for ``in-the-wild'' speaker verification. Hence, the larger test sets $Voxceleb1-E$ and $Voxceleb1-H$, derived from the entire Voxceleb1 dataset (including both the training set and test set), can verify the performance of the model accurately. It is worth mentioning that the testing set limits each test pair to include the speakers with the same nationality and gender, which requires the speaker verification model to learn more precise speaker identity features to achieve better performance. 

As shown in Table \ref{vox2}, SEEF-ALDR can reinforce the performance of the original models, with an average of 23.8\% improvement on EER. We find that the overall improvement of \cite{xie2019utterance} is less than \cite{chung2018voxceleb2}. We speculate the reason might be that NetVLAD (NV) and GhostVLAD (GV) proposed in \cite{xie2019utterance} can optimize the extraction of speaker embeddings, thus their EERs is already improved, much better compared with those of \cite{chung2018voxceleb2}. Therefore, it does not leave lots of optimization space for SEEF-ALDR, compared with \cite{chung2018voxceleb2}. Meanwhile, the improvement on the larger testing datasets $Voxceleb1-E$ and $Voxceleb1-H$ always falls behind that on a relatively small testing dataset Voxceleb1 for both \cite{xie2019utterance} and \cite{chung2018voxceleb2}. Last but not least, the best EER achieved by SEEF-ALDR as an optimization strategy over the baseline \cite{xie2019utterance} is 2.61\%, which is better than the results of most state-of-the-art research \cite{nagrani2017voxceleb,chung2018voxceleb2,xie2019utterance,cai2018exploring,hajibabaei2018unified,li2017deep,bhattacharya2017deep,okabe2018attentive,kim2019deep} (ranging from 2.85\% to 10.2\%).

\begin{figure}
  \centering
  \includegraphics[width=0.5\textwidth]{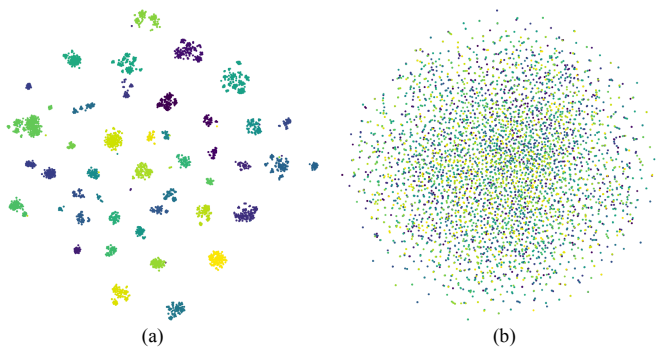} 
  \caption{T-SNE Visualization of the Decoupled Features based on the test set of Voxceleb1: (a) Features Extracted by $E_{p}$ (b) Features Extracted by $E_{e}$.}
  \label{compare1}
\end{figure}

\subsection{Performance of Disentangled Representation}
\begin{figure}
  \centering
  \includegraphics[width=0.5\textwidth]{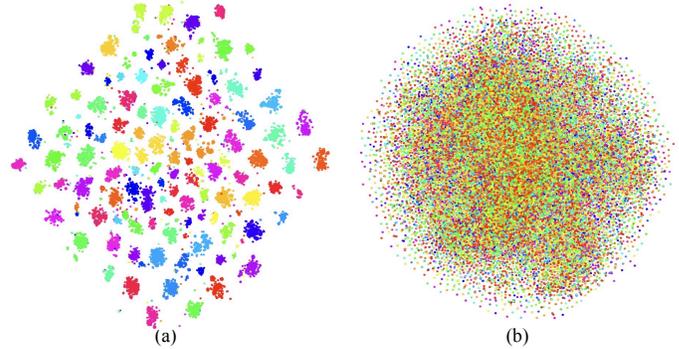} 
  \caption{T-SNE Visualization of the Decoupled Features based on the test set of Voxceleb2: (a) Features Extracted by $E_{p}$ (b) Features Extracted by $E_{e}$.}
  \label{compare2}
\end{figure}

The proposed framework SEEF-ALDR relies on the purifying encoder $E_{p}$ to extract speaker identity features and the eliminating encoder $E_{e}$ to extract identity-unrelated features. To demonstrate the difference between the two decoupled features, we use T-distributed Stochastic Neighbor Embedding (T-SNE) \cite{maaten2008visualizing} to reduce the dimension of the high-level features and visualize them. Figure \ref{compare1} shows the distribution of speaker identity representation learned by the purifying encoder and the eliminating encoder respectively, when SEEF-ALDR is trained on dataset Voxceleb1 based on model ResNet-34 \cite{cai2018exploring}. As shown in Figure \ref{compare1} (a), the identity of each speaker has a dense set of clustered features, and there are clear classification boundaries among the features of each identity. Hence, it demonstrates that $E_{p}$ can effectively extract identity-related information. In contrast, in Figure \ref{compare1} (b), the identity of each speaker is evenly distributed in the feature space, and the features of each identity overlap significantly with each other. Therefore, it shows that $E_{e}$ is capable of erasing the speaker identity information, thus constructing the identity-unrelated representation. 
Figure \ref{compare2} shows the distribution of speaker identity representation on a much larger testing dataset Voxceleb2 based on model ResNet-50 \cite{chung2018voxceleb2}, with 118 distinct identities, compared with 40 identities when using the testing dataset Voxceleb1 in Figure \ref{compare1}. We can still observe similar distribution as that of Figure \ref{compare1}, which indicates the proposed SEEF-ALDR is capable of decoupling identity related and unrelated features effectively even on the dataset with much more identities.

\subsection{Ablation Study}
To further evaluate the contribution of each module in SEEF-ALDR, a comprehensive ablation study has been conducted, with SEEF-ALDR built upon ResNet-34 used in \cite{chung2018voxceleb2}, that is, porting ResNet-34 into the framework of SEEF-ALDR. We use Voxceleb2 as the training dataset, Voxceleb1 as the testing dataset, and the same metrics, the detection cost function ($C_{det}$) and the Equal Error Rate (EER) as in the above experiments to evaluate the performance of speaker verification. 

SEEF-ALDR is mainly composed of three components: the purifying encoder, the eliminating encoder, and the reconstruction decoder. Among them, the eliminating encoder contains two adversarial losses: $L^{adv}_{s}$ (in Equations \ref{formula5}) and $L^{adv}_{e}$ (in Equations \ref{formula6}). Since the encoder is an essential component to extract speaker embeddings, which is then used to perform the speaker verification task, either the purifying encoder or the eliminating encoder in SEEF-ALDR cannot be removed or replaced. Then we gradually add other components on top of the encoder and train several variant models as shown in Table \ref{ablation}, e.g., the purifying encoder and reconstruction decoder ($E_{p}$ + $D_{r}$). Furthermore, it is possible that a randomly-chosen vector for the eliminating encoder will simply ignore the input spectrogram and output a random vector, which might still help train the adversarial classifier more or less. Therefore, to justify the contribution of our eliminating encoder, we replace it with a random vector and evaluate the performance of speaker verification accordingly, i.e., $E_{p}$ + $Random Vector$ + $D_{r}$ in Table \ref{ablation}.

Table \ref{ablation} shows the evaluation results of our ablation study. The performance of using our purifying encoder only is considered as the baseline performance for the following ablation experiments. A structure similar to an autoencoder that uses only the decoder and the encoder (i.e., $E_{p}$ + $D_{r}$) can improve the performance of speaker verification by 12.3\% over the baseline. However, such improvement is due to the capability of the autoencoder to extract embeddings from the input spectrogram, but still far away from that of our framework SEEF-ALDR. The eliminating encoder itself (i.e., $E_{e}$) gets very low accuracy in speaker verification, i.e., downgrading the baseline EER by 887.9\%, which also indicates that it is effective to extract identity-unrelated features, thus completely failing the speaker verification task. 

We also evaluate the impact of each adversarial loss individually by removing either $L^{adv}_{s}$ or $L^{adv}_{e}$, thus obtaining the performance of speaker verification on $E_{e}$ w/o $L^{adv}_{s}$ (drop by 504.\%) and $E_{e}$ w/o $L^{adv}_{e}$ (drop by 585.9\%). These results demonstrate that neither of the adversarial losses $L^{adv}_{s}$ nor $L^{adv}_{e}$ can accomplish the goal of eliminating identity-related information individually. They collaborate together and contribute to the training of an effective eliminating encoder. Furthermore, we find that the performance of replacing our eliminating encoder with random vectors ($E_{p}$ + $Random Vector$ + $D_{r}$) is almost equivalent to that of $E_{p}$ + $D_{r}$, i.e., 13.6\% increase vs 12.3\% increase, not able to significantly improve the performance of speaker verification. Overall, leveraging the contribution of the purifying encoder, the eliminating encoder, and the reconstruction decoder together, SEEF-ALDR can archive superior improvement of speaker verification, i.e., 38.8\% increase over the baseline. 


\begin{table}
\caption{Ablation Study for SEEF-ALDR. EER-IP represents the EER Improvement over the Baseline.}
\label{ablation}
\centering
\begin{tabular}{lccc}
\toprule
Branch  & EER (\%) & $C_{det}$  & EER-IP \\ \midrule
$E_{p}$                        & 5.04   & 0.543    & - \\ 
$E_{p}$ + $D_{r}$                     & 4.42   & 0.509    & 12.3\% \\\midrule
$E_{e}$                     & 49.79   & 0.999    & -887.9\% \\ 
$E_{e}$ w/o $L^{adv}_{s}$   & 30.46   & 0.999    & -504.4\% \\
$E_{e}$ w/o $L^{adv}_{e}$   & 34.57   & 0.999    & -585.9\% \\ \midrule
$E_{p}$ + $Random Vector$ + $D_{r}$             & 4.35   & 0.511    & 13.6\% \\
\textbf{SEEF-ALDR($E_{p}$ + $E_{e}$ + $D_{r}$)}           & \textbf{3.08}   & \textbf{0.334}    & \textbf{38.8\%} \\ \bottomrule
\end{tabular}


\end{table}

\section{Discussion}

\noindent\textbf{Novel identity impersonation attack based on SEEF-ALDR.}
Interestingly, a novel attack against speaker identity verification can be possible via the SEEF-ALDR framework based on voice conversion. As discussed above, the speaker identity feature from the purifying encoder determines the identity of the speaker, while the identity-unrelated feature from the eliminating encoder contains other semantic-rich information, such as language, emotion, environmental noise, etc. Suppose SEEF-ALDR learns speaker identity features and identity-unrelated features of Speaker A and Speaker B, respectively. Then two new speeches can be synthesized by the reconstruction decoder based on the cross-fusion the two decoupled features of them, e.g., identity features of Speaker A and identity-unrelated features of Speaker B, and vice versa. Hence, a flexible identity impersonation with the semantic-rich context can be performed, though the quality of the synthesized new speech depends on the optimization of the reconstruction process. We plan to optimize the objective function of the reconstruction decoder of SEEF-ALDR to produce a high-quality, semantic-rich speaker identity impersonation attack, and conduct a comprehensive analysis of existing speaker authentication mechanisms against such novel attack.

\noindent\textbf{Reinforcing sound event detection based on SEEF-ALDR.}
The proposed framework SEEF-ALDR also provides a potential solution to improve the performance of sound or audio event detection. Sound contains a lot of information about physical events that occur in our surrounding environment. Through sound event detection, it is possible to perceive the context of the sound, e.g., busy road, office, grocery, etc., and at the same time recognize the sources of the sound, e.g., car passing-by, door opening, footsteps, etc. Furthermore, sound event detection also has great potential to be used in a variety of safety-related scenarios, such as autonomous driving, intelligent monitoring systems, healthcare, etc. For instance, in the scenario of home security monitoring, sound event detection can analyze whether there is a stranger break-in through audio signal, to compensate for the blind area of surveillance cameras. 

However, in the real world, the audio signal usually contains sound from multiple sources simultaneously and is distorted by environmental noise. Therefore, it is nontrivial to accurately and efficiently extract the sound of interest from the information-rich audio to identify the event. Similar to speaker verification, existing works on sound event detection focus on optimizing the machine learning models to improve the detection performance, rather than decoupling the event-related and event-unrated features \cite{kumar2016audio,kao2018r}. Hence, the framework of SEEF-ALDR can directly help improve the performance of the audio event detection task. In particular, the eliminating encoder can be extended to remove the event-related features via adversarial learning, while the purifying encoder preserves most of the event-related information for classification. The decoder collects the output from both the purifying encoder and the eliminating encoder to reconstruct the spectrogram, which is made to resemble the original input spectrogram as much as possible. We plan to extend our SEEF-ALDR to the domain of sound event detection and make it a general approach for any task that benefits from feature decoupling of the original input.

\section{Conclusion}
Speaker verification has been used in various user authentication scenarios, so its dependability greatly impacts users' privacy and overall system security. In this paper, we propose SEEF-ALDR, a novel speaker embedding enhancement framework via adversarial learning based disentangled representation, to decouple the speaker identity features and the identity-unrelated ones from original speech. SEEF-ALDR is built based on twin networks integrating an autoencoder-like architecture and adversarial learning approach. The twin networks contain the speaker eliminating encoder and the speaker purifying encoder, with the former erasing speaker identity features and the latter extracting the speaker identity features. In this way, by combining adversarial supervision signal and reconstruction supervision signal, our framework can naturally learn complementary feature representations of the original speech, which means the fusion of the decoupled features is close enough to the original one. Experimental results demonstrate that, as an optimization strategy, SEEF-ALDR can construct more accurate speaker embeddings for existing speaker verification models to improve the performance of ``in-the-wild'' speaker verification with little effort.

\begin{acks}
This work was partially supported by the National Key Research and Development Program of China (2016QY04W0903), Beijing Municipal Science and Technology Project (Z191100007119010), Strategic Priority Research Program of Chinese Academy of Sciences (No. XDC02010900) and National Natural Science Foundation of China (NO.61772078). 
Any opinions, findings, and conclusions or recommendations expressed in this material are those of the authors and do not necessarily reflect the views of any funding agencies.
\end{acks}

\bibliographystyle{ACM-Reference-Format}
\bibliography{ref}


\end{document}